\documentclass[usenatbib,useAMS,a4paper]{mn2e}

\usepackage{txfonts}
\usepackage{epsfig}
\newcommand{\aap}{A\&A}
\newcommand{\araa}{AnnRevA\&A}
\newcommand{\mnras}{MNRAS}
\newcommand{\apj}{ApJ}
\newcommand{\apjl}{ApJL}
\newcommand{\apjs}{ApS}
\newcommand{\aj}{AJ}
\newcommand{\pasj}{PASJ}

\def\msun{{\rm M_{\odot}}}

\title [Metallicity, planet formation, and disc lifetimes]{Metallicity, planet formation, and disc lifetimes}
\author[]{B. Ercolano$^{1,2}$ and C.J. Clarke$^1$\\
$^1$Institute of Astronomy, Madingley Rd, Cambridge, CB3 0HA, UK \\
$^2$Department of Physics and Astronomy, University College London, WC1E 6BT, UK}

\date{Submitted:}

\pagerange{\pageref{firstpage}--\pageref{lastpage}} \pubyear{2003}

\begin{document}
\def\lta{\mathrel{\spose{\lower 3pt\hbox{$\mathchar"218$}}
     \raise 2.0pt\hbox{$\mathchar"13C$}}}
\def\gta{\mathrel{\spose{\lower 3pt\hbox{$\mathchar"218$}}
     \raise 2.0pt\hbox{$\mathchar"13E$}}}
\def\Msun{{\rm M}_\odot}
\def\msun{{\rm M}_\odot}
\def\Rsun{{\rm R}_\odot}
\def\Lsun{{\rm L}_\odot}
\def\19{GRS~1915+105}
\label{firstpage}
\maketitle

\begin{abstract}

The lifetime of protoplanetary discs is intimately linked to the
mechanism responsible for their dispersal. Since the formation of
planets within a disc must operate within the time frame of disc
dispersal it is crucial to establish what is the dominant process that
disperses the gaseous component of discs around young stars. 
Planet formation itself as well as photoevaporation by energetic
radiation from the central young stellar object have been 
proposed as plausible dispersal mechanisms. There is however still no
consensus as what the dominant process may be. In this paper we use
the different metallicity dependance of X-ray photoevaporation and
planet formation to discriminate between these two processes. 
We study the effects of metallicity, $Z$, on the dispersal timescale,
$t_{\rm phot}$, in the context of a photoevaporation model, by
means of detailed thermal calculations of a disc in hydrostatic
equilibrium irradiated by EUV and X-ray radiation from the central
source. Our models show $t_{\rm phot} \propto Z^{0.52}$ for a pure
photoevaporation model. By means of analytical estimates we derive
instead a much stronger {\it negative} power dependance on metallicity
of the disc lifetime for a dispersal model based on planet formation. 

A census of disc fractions in lower metallicity regions should
therefore be able to distinguish between the two models. A recent
study by Yasui et al. in low metallicity clusters of the extreme outer
Galaxy ([O/H]$\sim$-0.7dex and dust to gas ratio of $\sim$0.001)
provides preliminary observational evidence for shorter disc
lifetimes at lower metallicities, in agreement with the 
predictions of a pure photoevaporation model. While we do not
exclude that planet formation may indeed be the cause of some of the
observed discs with inner holes, these observational findings and the
models and analysis presented in this work are consistent with X-ray
photoevaporation as the dominant disc dispersal mechanism. 

We finally develop an analytical framework to study the effects of
metallicity dependent photoevaporation on the formation of gas giants
in the core accretion scenario. We show that accounting for this effect
strengthens the conclusion that planet formation is favoured at higher
metallicity. We find however that the metallicity dependance of
photoevaporation only plays a secondary role in this scenario, with
the strongest effect being the positive correlation between the rate
of core formation and the density of solids in the disc. 

\end{abstract}


\section{Introduction}

  It is not yet established what is the dominant process that disperses
the gaseous component of the discs around young stars. What is clear
observationally is that evidence of gas accretion onto stars, together with
infrared/submm  diagnostics associated with small dust grains entrained in the
gas, disappear in young stars at an average age of a few Myr (Haisch, Lada \& Lada 2001).
This figure needs to be interpreted as an {\it average} figure since clearly
individual stars may lose their discs on timescales that differ from this
by at least a factor three (Armitage, Clarke \& Palla 2003).
It is equally clear, based on the relatively few systems observed
in a state of transition between disc possessing and discless status, that
the timescale for disc dispersal (specifically the timescale over which
regions of the disc are optically thin in the infrared) is much shorter
than the overall disc lifetime (Skrutskie et al. 1990, Kenyon \& Hartmann 1995,
Duvert et al. 2000). The fact that transition discs are relatively rare
(constituting around $10 \%$ of the population of young stars with discs)
is however an observational hindrance when it comes to establishing the
mechanism for disc clearing: transition discs appear to be a rather
heterogeneous class of objects and it is hard to classify their diversity when
the number of well studied objects in nearby star forming regions is still
in single figures. An important feature of many transition discs, 
however, is that their spectral energy distributions are best fit by
inner holes in the disc (these holes are not necessarily devoid of dust,
but there is a large contrast in surface density between the outer disc
and inner cavity).\footnote{This situation may not extend to discs around
lower mass stars, as there are recent claims that discs in M stars may
instead pass through an extended phase in which their dust content is
homogeneously depleted at all radii (Currie et al. 2009).}

 Although an obvious mechanism for the disappearance of  dust diagnostics
in discs is the simple coagulation of grains into entities that are
large compared with the wavelength of observation (Dullemond \&
Dominik 2005), this scenario does not of itself explain why, in
general, there is a correlation between the 
disappearance of dust and gas accretion diagnostics. Moreover, it is
not obvious why dust coagulation should create a well defined inner hole 
structure rather than homogeneous depletion at all radii (although
models combining grain growth with photophoresis - once the inner disc
is optically thin to visible light in the radial direction  - are
promising in this regard; Krauss et al. 2007). 

 The two leading disc dispersal mechanisms that satisfy these
 considerations are the formation of giant planets and
 photoevaporation. Both these mechanisms create an inner hole in the
 disc and also disrupt the accretion flow onto the star to some extent\footnote{
 More or less completely in the case of photoevaporation (Clarke et
 al 2001, Alexander, Clarke \& Pringle 2006b) and to a variable extent
 in the case of planet formation, depending on the mass of the planet
 (Lubow et al. 1999, Rice et al. 2003).}. 
In the case of photoevaporation,
inner hole creation is succeeded by the rapid clearing of the outer disc
thereafter. In the case of planet formation, clearing of the outer disc
requires further planet formation at larger radii, i.e. the relative
scarcity of transition discs argues that planet formation at one
radius is rather rapidly followed by a wave of successive planet formation
events in the outer disc ( Armitage \& Hansen  1999). Some
attempts have been made to classify 
whether individual transition discs are likely to be  generated
by planet formation or photoevaporation (Najita et al. 2007, Alexander
\& Armitage 2007,  Alexander 2008, Cieza et al. 2008, Kim et al. 2009) based
on estimates of the accretion rate and disc mass. These analyses are
however based on extreme ultraviolet (EUV) photoevaporation models
(Hollenbach et al. 1994, Clarke et al. 2001, Alexander, Clarke \&
Pringle 2006a,b) which predict low accretion rates and disc masses in
transition objects, in contrast to many of the observed systems. More
recent models based on X-ray photoevaporation (Ercolano et al
2008, 2009, Owen et al. 2009) predict higher accretion rates and disc
masses during transition and thus somewhat muddy the observational
distinction between the two classes of disc dispersal
mechanism. Nevertheless it would appear clear that 
no single mechanism can explain all observed inner hole systems: 
for example, photoevaporation is incompatible with the predominantly
high accretion rates in the sample of Kim et al. (2009) whereas a
planet is perhaps an unlikely candidate for many of the objects
contained in Cieza et al. (2008), which have very large holes and low
accretion rates.

\begin{figure*}
\begin{center}
\begin{minipage}{20cm}
\includegraphics[width=9.0cm]{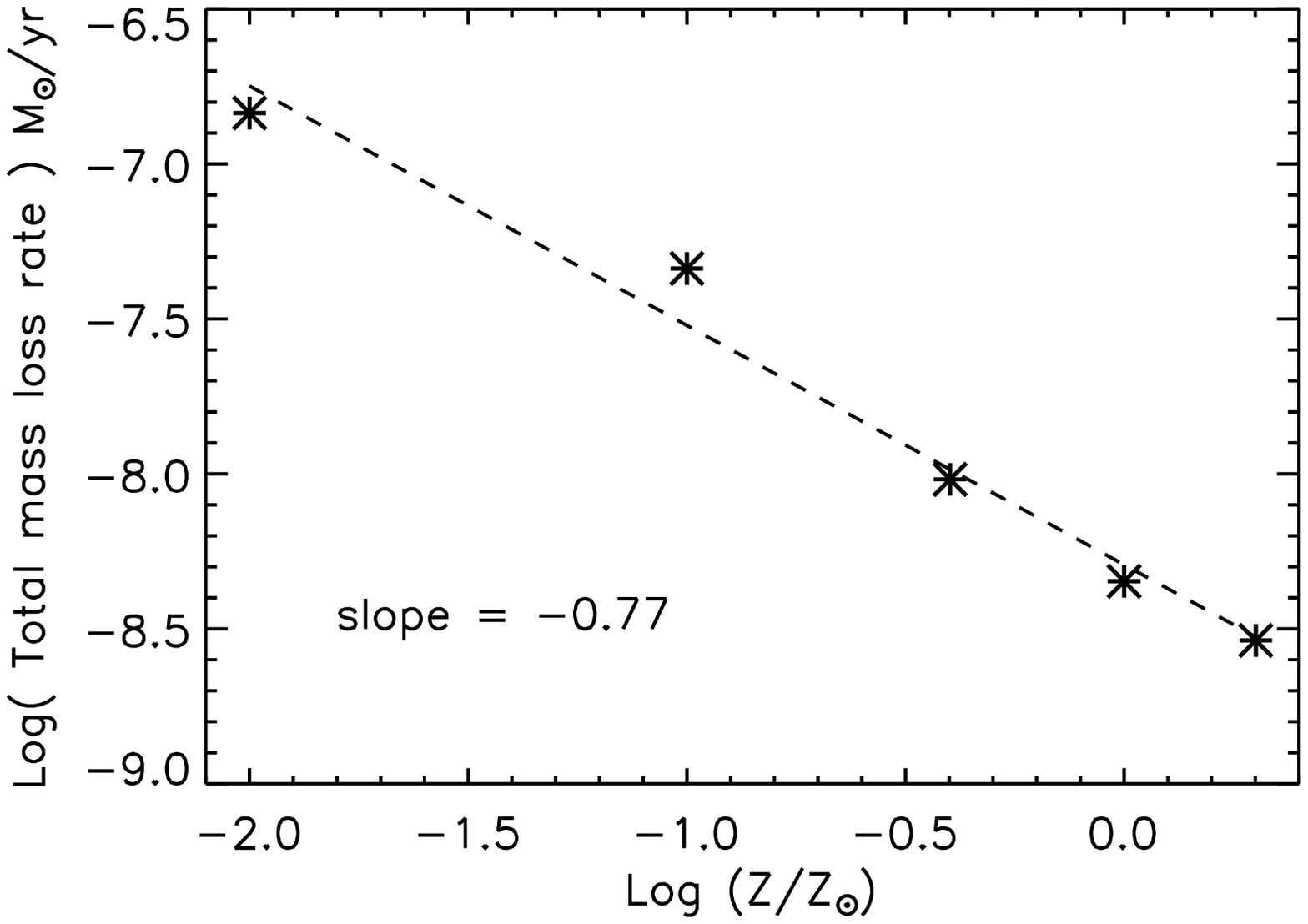}
\includegraphics[width=9.0cm]{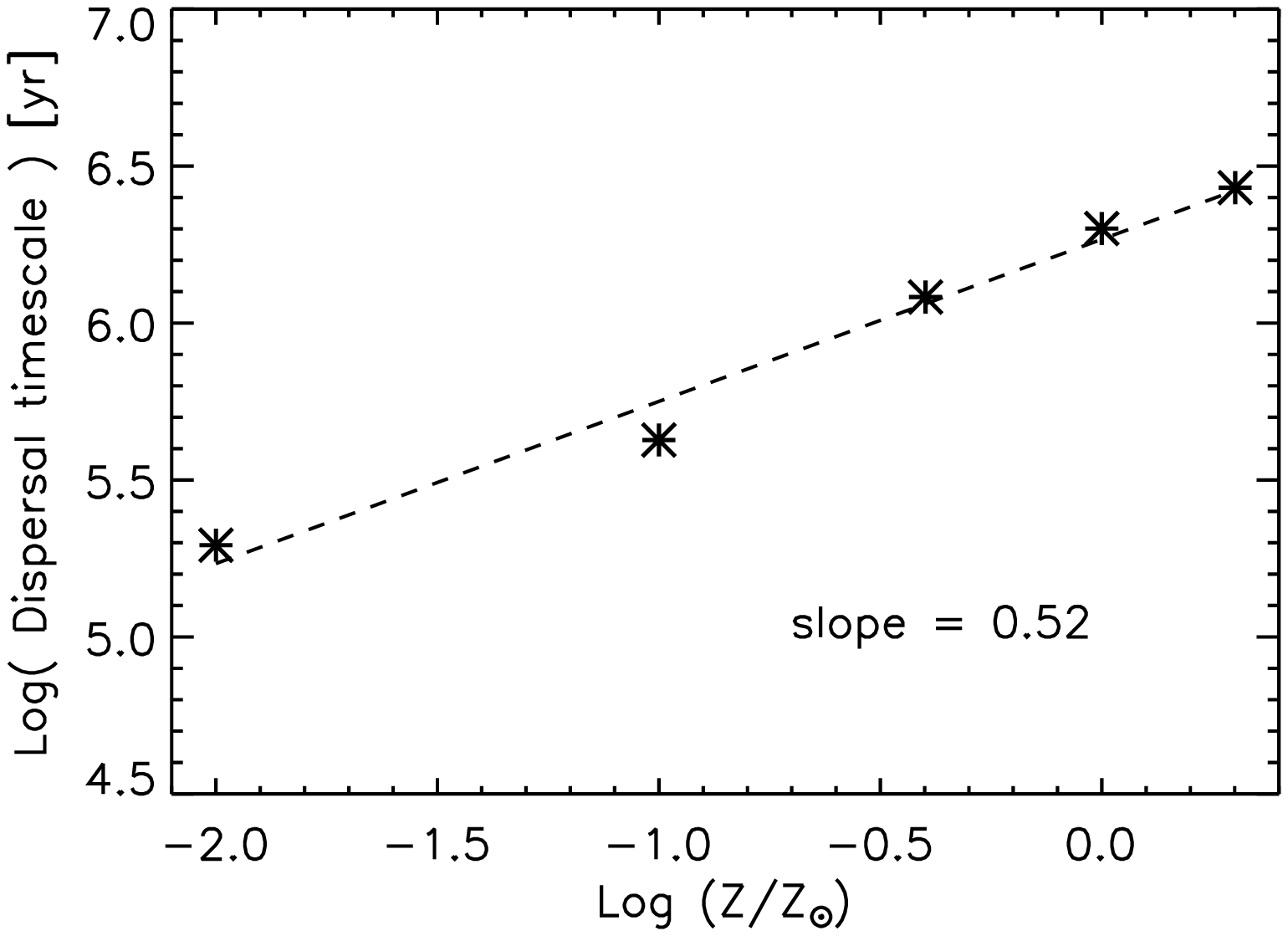}
\end{minipage}
\caption[]{The metallicity dependance of mass loss rates (left panel)
  and dispersal timescales (right panel) of protoplanetary discs 
  photoevaporated by X-ray and EUV from the central star. {\it Right} 
Model data points (asterisks) obtained from our photoionisation
modeling; the dashed line is a least square fit to these
data. {\it Left:} The asterisks show the photoevapoaration timescales
obtained substituting the points on the right panel into equation A10
and normalising to 2~Myr for solar metallicities. The dashed line
shows the relation given in equation 4 for p~=~1.}
\label{f:mods}
\end{center}
\end{figure*}

 At this point it is perhaps worth noting - despite the obvious
interest of observed transition objects - that they do not {\it
necessarily} hold the key to the process that disperses the
{\it majority} of discs around young stars. This is because
objects observed in transition can in principle be a mixture
of stars that spend a significant time in this state (and
must therefore be a minority of all young stars on statistical
grounds) and a general population of `typical' young stars
that must pass through such a transition quickly. (The argument
that some fairly abrupt end to a disc's lifetime is required is
well illustrated by considering what would happen if a disc merely
continued to accrete onto the central star as a result of
viscous evolution - the observed accretion rates, ages and disc
masses of T Tauri stars would imply that integrated forward in time
they would remain optically thick for $> 100$Myr and then would
linger for a comparable period as partially cleared systems).
Therefore  - regardless of what process explains the bulk 
of observed transition discs - we also need to decide what is
the mechanism that terminates the lifetime of the majority
of disc bearing young stars. 

  In this paper we therefore take a different approach to
  discriminating between planet formation and photoevaporation as the
  prime disc dispersal mechanism, by instead considering the {\it
    metallicity dependence} of these two processes. Although most
  studies of circumstellar discs have currently been undertaken in
  regions with a rather low range of metallicities, the advent
  of the Herschel Space Observatory and later the James Webb Space
  Telescope raises the prospect of being able to conduct disc censuses
  in more distant regions whose metallicity differs considerably from
  that of star forming regions in the solar neighbourhood. Such
  studies are already possible in the Extreme Outer Galaxy (EOG) with
  current instrumentation as demonstrated by the recent work of Yasui
  et al. (2009) which we will discuss in more detail later. 
  Our focus here will not be on the nature of the (in any case rare)
  transition objects, but will instead centre on how the {\it mean disc lifetime
    is expected to vary with metallicity in these two scenarios}. To this
  end we combine simple models for disc viscous evolution  with
  (metallicity dependent) photoevaporation and also with semi-analytic
  prescriptions for planet formation (Section 2). We will show that
  in the photoevaporation model the disc lifetime is a mildly increasing
  function of metallicity, resulting from the rather higher photoevaporation
  rates in the case of low metallicity gas for which opacities are
  lower and line cooling less efficient. On the other hand, the
  disc lifetime is a strongly decreasing function of metallicity in
  the case that disc clearing is initiated by planet formation. This
  is simply because at  higher metallicity the assumed higher surface
  density of planetesimals in the disc encourages the more rapid
  formation of a gas giant planet, specifically because it accelerates
  the formation of a rocky core of the critical mass required to
  initiate the accretion of a gaseous envelope. This is borne out, for
  example, by the hydrodynamical models of gas giant formation of
  Pollack et al. (1996) and Hubickyj et al. (2005)\footnote{Note that
    this conclusion is based on an assumed linear relationship between
    the density of planetesimals and the metallicity; this conclusion
    would be only strengthened if one takes into account the recent
    suggestion of  Anders et al 2009  that the efficiency of planetesimal
    formation by the streaming instability should increase steeply at
    higher metallicity.}. 
  This positive link between metallicity and 
  planet formation has been noted by a number of authors (e.g. Ida \&
  Lin 2004b)
  as a possible explanation of the observed increase in the frequency
  of giant planets as a function of metallicity (Santos, Israelian \&
  Mayor 2000, 2001, 2004; Gonzalez et al. 2001; 
  Sadakane et al. 2002; Heiter \& Luck 2003; Laws et al. 2003; Fischer
  \& Valenti 2005; Santos et al. 2002 and references therein) but has not
been previously linked to the issue of disc lifetimes. In Section 3, we
revisit the connection between the frequency of giant planets and 
metallicity in the case of a hybrid model where planet formation
takes place in the context of a nebula that is subject to metallicity
dependent photoevaporation of the disc gas. Since photoevaporation
is more effective at low metallicities and thus reduces the disc lifetime,
this strengthens the conclusion that planet formation is favoured
at high metallicity. Nevertheless, the  additional ingredient 
of metallicity dependent photoevaporation is found
to be a second order effect in determining the positive correlation
between planet frequency and metallicity. We present our conclusions in Section 4. 

\section{The dependence of disc lifetime on metallicity}

\subsection{Lifetime against photoevaporation}

 We consider the case where the gas in the disc (initially with mass
$M_{go}$) undergoes viscous accretion onto the central star. As a result
of this secular evolution, the disc radius grows while the disc mass
and the accretion rate onto the star decline with time. We here parameterise
this process according to the similarity solutions that pertain in the
case that the kinematic viscosity in the disc is a simple power law function
of radius ($\nu \propto r^p$; Lynden-Bell \& Pringle 1974, Hartmann et al 1998).
This is not likely to be a good description in detail since even in a simple
$\alpha$ disc model the viscosity is a function of both surface density and
radius, depending  on the dominant local opacity source; yet further
complications are introduced when more realistic prescriptions for angular
momentum transport -- involving the action of the magnetorotational
instability or of self-gravitating torques are included
(e.g. Armitage et al. 2001, Clarke 2009,  Rice  \& Armitage 2009, Zhu et
al 2009, Cossins et al. 2009). 
The prescription we use here however has the advantage of providing a
simple analytic form such that the value of $p$ is linked to the power
law index of the surface density profile (i.e. over much of the radial
range of the similarity solution, the  disc's surface density is a power law
$\Sigma \propto r^{-p}$). Our choice of $p$ (which fixes the power
law exponents for the evolution of disc quantities) is therefore 
observationally motivated by fits to the infrared spectral/submm
energy distributions of young stars, which suggest $p \sim 1$ (Andrews et al
2009, Kitamura et al 2002, Isella et al 2009)

\begin{figure*}
\begin{center}
\begin{minipage}{20cm}
\includegraphics[width=9.0cm]{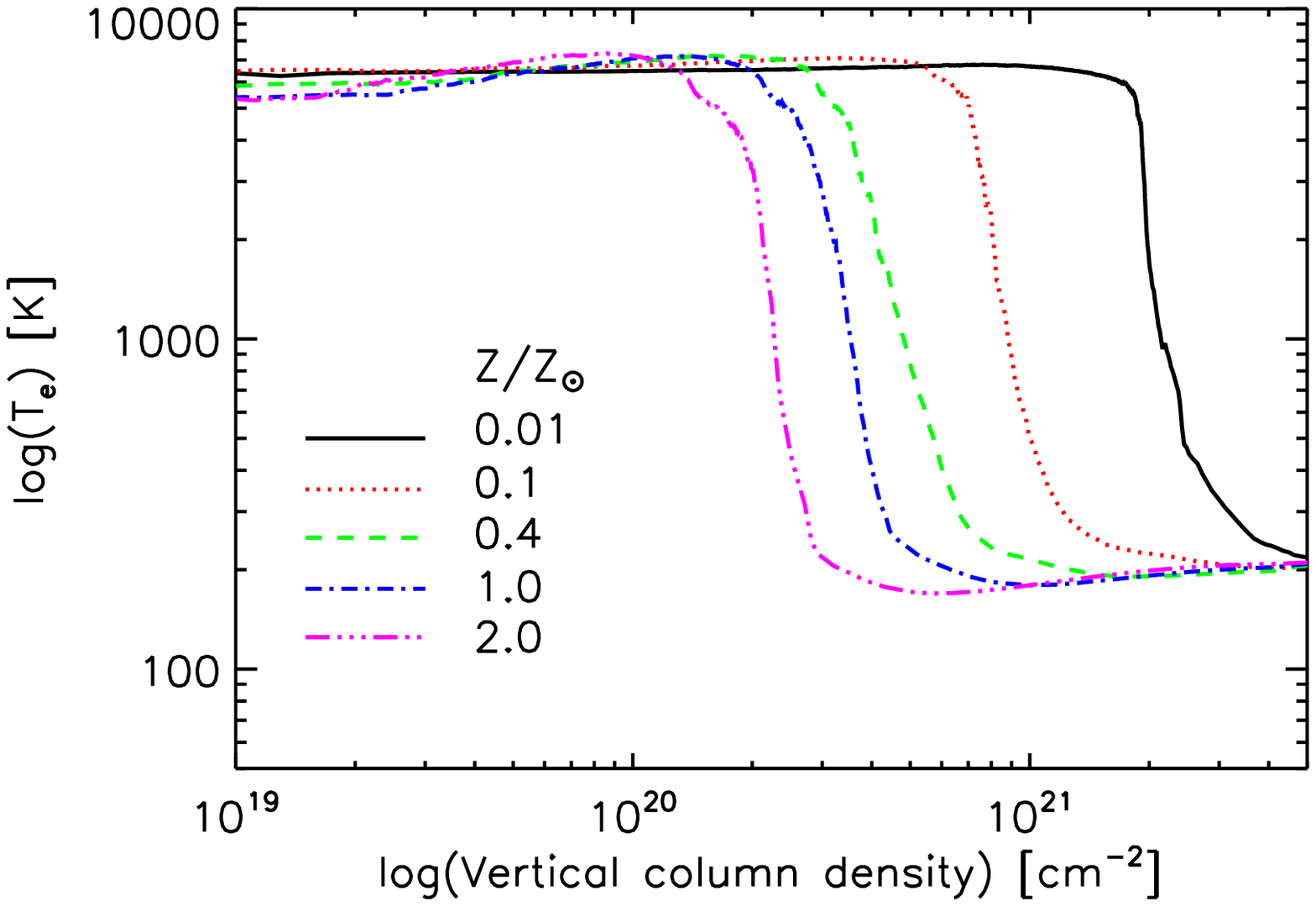}
\includegraphics[width=9.0cm]{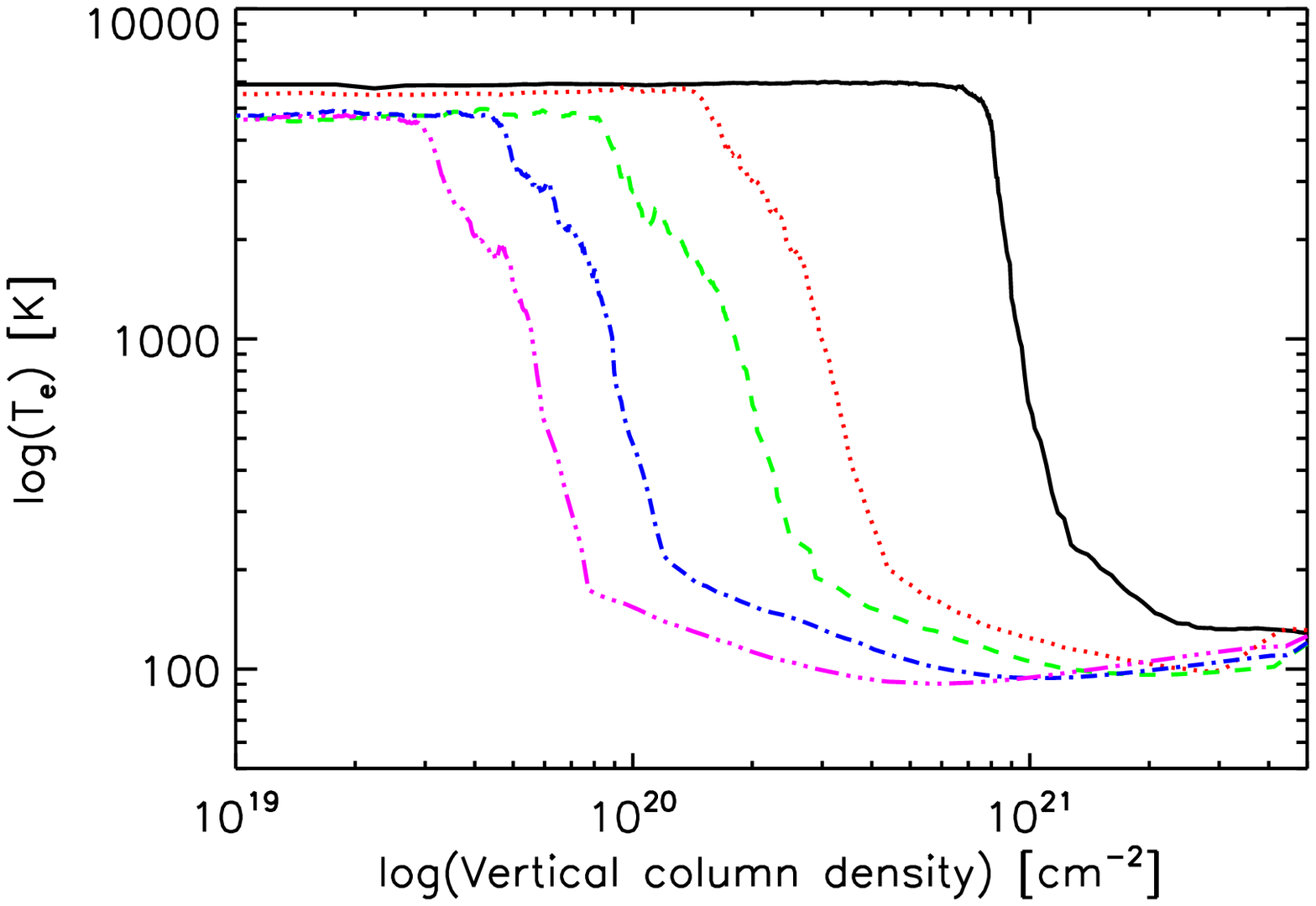}
\end{minipage}
\caption[]{Gas temperatures distribution at a radial distance of 5~AU
  (left panel) and 20~AU (right panel).}
\label{f:temps}
\end{center}
\end{figure*}

 The Appendix contains a heuristic derivation of the asymptotic
form of the similarity solutions; according to equation \ref{mdott}
the accretion rate declines as:

\begin{equation}
\label{mdott}
\dot M \propto  M_{g0} r_{d0}^{1/2} t^{(-5+2p)/(4-2p)}
\end{equation}

where $r_{d0}$ is the initial disc scaling radius. Such a power law
decline obviously does not enable one to define a disc lifetime, since
the timescale on which disc properties change by order unity is always
of order the present age. We therefore need to introduce a further condition
for disc dispersal. In the case that we invoke (metallicity dependent)
photoevaporation, at a constant rate $\dot M = \dot M_W(Z)$, then
the disc disperses rapidly at the point that the accretion rate
through the disc falls to $\dot M_W(Z)$\footnote{Here we are ignoring
  the fact that, as shown by Owen et al. (2009) the rapid disc
  dispersal begins once the accretion rate has fallen to
  roughly ten times below the photoevaporation rate, causing the
  system to experience a period of photoevaporation starved accretion
  (Drake et al. 2009) prior to the fast dispersal phase. These
  consideration however do not affect the proportionality relations
  and therefore the conclusions of this work remain unchanged.}. Thus the
disc lifetime against photoevaporation can be written:

\begin{equation}
\label{tphot}
t_{phot} \propto \dot M_W(Z)^{(4-2p)/(-5+2p)} M_{g0}^{(4-2p)/(5-2p)} r_{d0}^{(2-p)/(5-2p)}
\end{equation}

 At this point it becomes necessary to specify the source of the
photoevaporative flow. Until very recently it has been widely
assumed that such a flow is mainly driven by the extreme ultraviolet
(EUV) photons from the central star. However
we have recently shown that
soft (0.1-1keV) X-ray irradiation
can drive  powerful winds with typical mass loss rates of
10$^{-9}$-10$^{-8}$ M$_{\odot}$/yr (Ercolano et al., 2008b, 2009 ECD09,
Owen et al., 2009). In the hydrostatic equilibrium
 model of ECD09, X-rays from the
central pre-main-sequence star  ionise and heat the gas in the disc
atmosphere; the photoevaporation rate is estimated by assuming
that  gas that becomes hotter than the local escape temperature
becomes unbound and escapes at the sound speed. In reality, pressure
gradients within the disc are such that a photoevaporative flow can
in fact be initiated subsonically in deeper layers where the temperature 
is less than the
escape speed. This produces 
even larger winds (Owen et
al. 2009) which exceed photoevaporation rates from EUV radiation
by two orders of magnitude.
In fact, for strong X-ray emitters (i.e. with $L_X \sim 10^{30}$ erg s$^{-1}$),
these large  flow rates
are similar to the median  accretion rates
measured in T Tauri stars (Natta et al. 2006).  
When one also takes  account of the
dispersion in X-ray luminosities among young stars (Preibisch et al. 2005,
Albacete-Colombo et al. 2007) and the
associated range of expected photoevaporation rates, it then becomes
plausible that - in the absence of competing disc dispersal mechanisms -
the lifetime of gas discs in T Tauri stars is terminated by
X-ray powered photoevaporation. 

 With this in mind, we therefore use the
metallicity dependence of the photoevaporation rate derived from
our X-ray photoevaporation models. These employ
the modeling strategy described by Ercolano et
al. (2008b) and ECD09 to obtain the
temperature structure and photoevaporation rates for gaseous discs
with a range of metallicities, spanning from 0.01 solar to twice
solar. Briefly, this involves the coupling of the 3D photoionisation
and dust radiative transfer code MOCASSIN (Ercolano et al. 2003,
2005, 2006, 2008a) with an iterative solution of the  hydrostatic equilibrium
structure of the irradiated disc.
Our initial
guess at the density distribution is the disc model of d'Alessio et
al (2001) for a 4000~K 0.7~M$_{\odot}$ pre-main sequence star
surrounded by an optically thick flared circumstellar disc with an
outer radius of 500~AU and a total mass of 0.027~M$_{\odot}$. This
structure is irradiated by an unscreened EUV+X-ray spectrum (see
description in ECD09). 
We refer
to ECD09 for further details including the dust model and the
irradiating spectrum. Our solar abundance set is that of Asplund et
al. (2005) depleted according to Savage \& Sembach (1996). The dust to
gas ratio for the solar abundance case was set to 6.5$\times$10$^{-4}$ (see
discussion in ECD09). For models of different metallicity, the dust to
gas ratio and the abundances of metals were multiplied by
(Z/Z$_{\odot}$). As noted above, we estimate the photoevaporation
rates by assuming that a sonic flow is launched from regions where the
temperature exceeds the local escape temperature. Our preliminary
hydrodynamic calculations (Owen et al. 2009) suggest that this may
under-estimate the total 
mass flow rates, although we do not expect this to have a large effect
on the relative scaling with metallicity.

The left panel of Figure~\ref{f:mods} shows the dependance of the
total photoevaporation rates on metallicity, \.M$_{\rm W}(Z)$, which
can be approximated by a power-law of index -0.77. The model data
points shown in the left panel of this figure were obtained using the approach
described above, which involved rerunning model FS0H2Lx1 from the set
described by ECD09 for the appropriate metallicities and dust-to-gas
ratios. The corresponding photoevapoaration timescales, shown in the
right panel, were then obtained by substituting these numerical values of
\.M$_{\rm  W}(Z)$ in equation A10. The resulting timescales were then 
normalised such that a lifetime of 2~Myr is obtained at solar
metallicities. 

The increasing  X-ray photoevaporation rates at lower metallicity can
be readily understood by looking at the gas temperature distributions
in the disc for the different metallicity cases that are shown in
Figure~\ref{f:temps}. The reduced extinction in the low metallicity
cases allows high density gas at larger columns to be ionised and heated to
temperatures sufficiently high for the gas to be entrained into a
photoevaporative flow. Other metallicity-dependent effects are also
present and work in the same direction (i.e. lower metallicity =
higher gas temperatures), but they play a secondary role. These
effects are listed here for completeness: a low metallicity implies
a reduced cooling by fine structure lines of ions and 
neutrals (such as [O~{\sc i}] and [C~{\sc ii}], e.g. Ercolano et
al. 2008b), which leads to higher temperatures in the X-ray heated
disc atmosphere, as particularly evident in the right panel of
Figure~\ref{f:temps}. Also a lower dust to gas ratio means
that the gas experiences a reduced competition from the dust grains
for the absorption of the energetic photons and that the dust-gas
collisional cooling term, which may provide a significant contribution
at the base of the photoevaporative envelope, is reduced at lower
metallicities. We emphasise however that by far the dominant
mechanism here is the reduced opacity which increases the penetration
column of the ionising radiation. It is worth noting at this
point that such a large effect is not expected to occur for gas
ionised by EUV only. X-ray photons are mainly absorbed by the inner shells of
the more abundant heavy elements in the gas and dust (e.g. oxygen,
carbon etc.) while for EUV photons by far the largest source of
opacity is hydrogen, which is not affected by changes in
metallicity (although a reduction in the dust-to-gas ratio would
still reduce the total EUV opacity). We also note that a
photoevaporation process based on far-ultraviolet (FUV) irradiation,
would also follow a different dependance on metallicity (Gorti \&
Hollenbach 2009). Therefore {\it the predicted metallicity dependance
  shown here pertains only to X-ray photoevaporated discs}. 

 When this metallicity dependent wind mass loss rate is combined with
equation \ref{tphot} one obtains the result that

\begin{equation}
\label{tphot2}
t_{phot} \propto Z^{0.77(4-2p)/(5-2p)} M_{g0}^{(4-2p)/(5-2p)} 
r_{d0}^{(2-p)/(5-2p)}
\end{equation}

Since we do not expect the distribution of initial disc gas
properties ($M_{g0}$ and $r_{d0}$) to depend on metallicity, we thus
deduce that the distribution of disc lifetimes in regions of 
different metallicity should scale with a Z dependent factor, i.e.

\begin{equation}
\label{tphot3}
t_{phot} \propto Z^{0.77(4-2p)/(5-2p)}
\end{equation}

  This Z dependence is rather weak, with a power law exponent of
$0.52$ (for $p=1$) and $0.38$ (for $p=1.5$).   
The right hand panel of Figure \ref{f:mods} is an illustrative example of
the dependence on metallicity of the mean disc lifetime of a population
of discs (with $p=1$) that is normalised such that  the mean disc lifetime for 
discs of  solar metallicity is $2$Myr. We emphasise that this normalisation
is set by our assumptions about the disc's secular evolution timescale,
as controlled by its equivalent viscous `$\alpha$' parameter (Shakura
\& Sunyaev 1973). We can readily adjust this quantity within a plausible
range in order to reproduce an observationally reasonable mean disc
lifetime.  An increase in metallicity from solar to twice solar causes the disc
lifetime to increase from $\sim$2 to $\sim$3.1~Myr, and a decrease
from solar to -0.7 dex solar, as in Cloud 2 in the EOG observations of
Yasui et al. 2009, produces a decrease in the disc lifetime from
$\sim$2 to $\sim$0.7~Myr, which is roughly consistent with the
observed disc fractions. 

\subsection{Disc lifetime against gas giant planet formation}

We now assume that disc dispersal is instead initiated by the formation
of the first gas giant planet, as is a common interpretation of observed 
transition discs (Rice et al. 2003, Setiawan et al. 2007).  According to the core accretion
model, the formation of a gas giant involves first the accumulation of a
solid core of critical mass (of order $10-20$ earth masses) followed by the
hydrodynamic accretion of gas from the surrounding disc (Pollack
et al. 1996). This latter occurs
on the Kelvin Helmholtz timescale of the planet, which is $\sim$ a Myr
for a $20$ earth mass planet and decreases strongly with increasing
planet mass. For the purpose of our analytic estimates (following
Ida \& Lin 2004a) we therefore assume that the limiting timescale in the
creation of a gas giant planet  (i.e. a roughly Jupiter mass planet which
can clear a gap in the
disc) 
is the time required to accumulate a core containing a critical  mass of
solids. In fact hydrodynamic calculations of planet formation paint
a more complex picture in which the majority of the formation process
is instead spent in a phase (`Phase 2';Hubickyj et al. 2005) where the
planet accretes a mixture 
of gas and planetesimals and where, depending on the properties of the
background nebula, gas giants can form, albeit more slowly, with a
core mass somewhat below the critical value mentioned above (we discuss
below how such considerations affect our conclusions).

 Proceeding for now with our simple analytic estimate
 we evaluate the 
timescale needed to achieve a fixed critical core mass (together
with  its scaling with metallicity),
by using  the analytic expression for the growth of solid core mass
($M_c$)  given in Ida \& Lin (2004a, IL04):

\begin{equation}
\label{mct}
{{dM_c}\over{dt}} \propto \Sigma_d(r) r^{-3/5} M_c^{2/3} \Sigma_g^{2/5}
\end{equation}
 
 where $\Sigma_d$ is the surface density of rocky planetesimals,
and $\Sigma_g$ is the instantaneous local value of the gas surface
density. We follow IL04 by assuming
that the planetesimal  distribution is decoupled from the evolutionary processes that control the evolution
of $\Sigma_g$. We thus assume that a fixed fraction of the initial solid content of the
disc forms a planetesimal
disc whose radial profile follows the initial  radial profile of the gas
and that $\Sigma_d$ is then constant in time (we thus neglect the depletion in $\Sigma_d$ due to the accretion of planetesimals during core growth; 
this is acceptable to first order
provided that the core is not close to achieving its isolation mass, an
issue to which we return below.) 

  We however (see Appendix) differ from IL04 in that we use the
viscous similarity solutions to determine the time dependence of
$\Sigma_g$ rather than adopting an ad hoc exponential
reduction in the gas surface
density. Solving equation \ref{mct2} and requiring that the core mass 
attains a fixed value for runaway gas accretion (see Appendix) then
implies a formation timescale:

\begin{equation}
\label{tform_p}
t_{form} \propto Z^{-5(2-p)/(5-3p)} M_{g0}^{-7(2-p)/(5-3p)} r_{d0}^{(2-p)(9-5p)/(5-3p)} r^{(2-p)(7p+3)/(5-3p)}
\end{equation} 

Since this timescale is an increasing function of radius, it follows
that planet formation is first favoured at small radius. However
this effect is counteracted by two effects that favour core growth
at larger radius. Firstly, the sticking efficiency of planetesimals
increases once the radius at which ice sublimes ($a_{ice}$) is
exceeded (Hayashi 1981, Pollack et al 1994). Secondly, it is necessary that a critical core mass can form without 
consuming all the planetesimals in its `feeding zone'.
This latter is often assumed to be an annular region
around the core with width related to the Hill radius of the core (i.e. of
width that scales linearly with orbital radius).
Thus the radius ($a_{tg}$) at which a core of critical mass just
consumes all the planetesimals in its feeding zone obeys a scaling of
the form\footnote{If, as
discussed in Ida \& Lin (2004), one relaxes this assumption so
that the core can feed from the entire stock of planetesimals interior
to its orbit then one also obtains a scaling of the same form.}

\begin{equation}
\label{atg}
\Sigma_d a_{tg}^2 = {\rm{constant}}
\end{equation}

i.e.
\begin{equation}
\label{atg2_p}
a_{tg} \propto Z^{-1/(2-p)} r_{d0} M_{g0}^{-1/(2-p)}
\end{equation}

  Therefore if $a_{ice} > a_{tg}$ (i.e. if the minimum radius for planet formation is set by ice sublimation
rather than feeding zone considerations) we have:

\begin{equation}
\label{tformmin1_p}
t_{form_{min}} \propto \bigl( Z^{-5} M_{g0}^{-7} r_{d0}^{(9-5p)} \bigr)^{(2-p)/(5-3p)}
\end{equation}

Alternatively, if $a_{tg}>a_{ice}$ we substitute \ref{atg2_p} into \ref{tform_p} and obtain

\begin{equation}
\label{tformmin2_p}
t_{form_{min}} \propto \bigl( Z^{(-13-2p)} M_{g0}^{-17} r_{d0}^{(4-2p)(6+p)} \bigr)^{1/(5-3p)}
\end{equation}

 We thus see that these planet formation timescales are highly
Z dependent, in the sense that planet formation is faster in discs
with higher metallicity. For $p=1$ and $p=1.5$ equation \ref{tformmin1_p}
implies that the planet formation timescale scales either with
$Z^{-2.5}$ or $Z^{-5}$, while \ref{tformmin2_p} implies scaling either
with $Z^{-7.5}$ or $Z^{-32}$!"\footnote{ We also note that
the planet formation timescale is highly dependent on the initial disc
mass, implying that planet formation should only be possible in systems
belonging to the upper centiles of the disc mass distribution. When the
disc mass threshold implied by the Ida \& Lin models is combined with
observations of the disc mass distribution by Andrews \& Williams (2005), the
predicted planet frequency is consistent with those found in radial
velocity surveys at solar metallicity: see discussion in Wyatt, Clarke \&
Greaves (2007).}.

  This extreme level of Z dependence is almost certainly an artifact of the
fact that we have assumed that the timescale for giant planet formation
is completely controlled by the timescale for solid core growth. It is
also worth mentioning that this very steep scaling stems from the fact that
the prescription for core growth (Ida \& Lin 2004a; equation~5 above)
involves a dependence on gas density; physically, this
accounts for the fact that gas drag reduces the velocity dispersion
of the planetesimals and hence increases the core's accretion
cross section due to gravitational focusing (Kokubo \& Ida 2002). We assume that the
disc gas surface density declines as a power law  
 due to  viscous
evolution and thus the core
grows  as a low power of time. Thus  the time required to achieve
a given core mass is a strong function of the normalisation of the solid
mass. This is not necessarily incorrect and is in this
respect  probably
more realistic than models of core accretion which adopt a temporally
constant background gas surface density (to date such models either
keep this quantity constant or introduce a  linear decline to
zero over $1-3$ Myr; Lissauer et al. 2009). 

  We now turn to the results of hydrodynamic models of planet formation
and see to what degree the limited range of available models in fact
support our estimates above.  These models demonstrate two effects:
increasing the surface density of solids decreases the timescale for
planet formation, in qualitative agreement with the arguments above.
Another factor in the opposite direction is that at higher
metallicity the opacity of the accreting gaseous envelope is increased;
this slows the contraction of the envelope and thus slows down gas
accretion. (Note that our simple model above only considers the time needed 
to acquire a critical solid core mass and does not consider the subsequent
time spent in acquiring a similar mass in gas which is considerable
in the hydrodynamic calculations; Hubickyj et al. 2005). Nevertheless, it
is found that the acceleration of planet formation due to enhanced
planetesimal surface density is a much stronger effect than the retarding
effect of increased opacity. For example, Hubickyj et al. found that
the time to form a gas giant increases by around a factor three if
the dust opacity is boosted by a factor fifty (see also Ayliffe \&
Bate 2009); on the other hand a modest
increase of the planetesimals surface density by a factor $1.6$ reduces
the formation timescale by a factor six. Thus the simulations support
our qualitative conclusion that the net effect of increasing the
metallicity is to strongly decrease the formation timescale of
giant planets 

 More detailed comparison is not really warranted given the fact that
the choice of simulations performed was motivated by different
scientific questions than the one we ask here. We are interested
in the time required to form a planet at given metallicity
{\it anywhere} in the disc. Since we have imposed a hard requirement
that the growing protoplanet should achieve a fixed core mass, we
exclude planet formation in inner regions of the disc where the isolation
mass is less than this critical value; in our prescription we first
form a planet at a larger radius where the formation timescale is longer.
Evidently this slows down planet formation at low metallicity. The
hydrodynamical calculations are instead motivated by creating Jupiter
at a particular radial location. In low metallicity runs planet formation
takes longer because the forming protoplanet has to `make do' with a lower
core mass and this then greatly increases the timescale for accreting
a comparable mass of gas. In a real disc both these effects would come into
play and it is not  clear whether, under these circumstances, the first
planet to form is at larger radius (where the core mass is higher but the
core accumulation timescale is longer) or at smaller radius where the 
bottleneck is instead the slow accumulation of gas onto a smaller mass
rock core. What is important to our discussion is that both effects are
in the same direction and imply a strong (negative) dependence of
planet formation timescale on planetesimal surface density (and thus
implicitly metallicity).

If the majority of protoplanetary discs dispersed due to planet
formation then a disc census in a lower metallicity region, such as
Cloud 2 in the EOG, would yield much longer disc lifetimes than those
derived in the solar neighbourhood. This is in contradiction with the recent
observations of Yasui et al. (2009) who find disc lifetimes shorter
than 1~Myr, compared to the few Myr found in the solar neighbourhood
(e.g. Haisch et al. 2001).

\section{The metallicity dependence of planet formation in discs
  subject to photoevaporation} 

As the list of exoplanets discovered in the solar neighbourhood continues 
to grow, statistical significance is lent to the observation that solar
type stars
hosting giant
planets  are on average more metal-rich than field stars
(Santos, Israelian \& Mayor 2000, 2001, 2004; Gonzalez et al. 2001;
Sadakane et al. 2002; Heiter \& Luck 2003; Laws et al. 2003; Fischer
\& Valenti 2005; Santos et al. 2002 and references  
therein). While the origin of this metallicity excess has been the object
of much debate in the literature, with the two main scenarios being
``primordial'' or ``external'' enrichment, there is a growing
consensus that  
metal rich gaseous discs are a  more
favourable environment for the formation of the rocky cores required by
core accretion models  (e.g. Ida \& Lin, 2004a).   

 The  phenomenological simulations of gas giant planet
formation of IL04 (whose assumptions imply, as we showed above,
a strong reduction
in planet formation timescale with metallicity)  demonstrated  a statistical preference 
for planet  formation and survival in discs with a high
ratio of solids to gas. If one ignores the possibility of stellar
surface contamination then the solid to gas ratio simply scales with
the metallicity of the central star. 
Such simulations therefore reproduce the observed  positive correlation between the incidence
of planets and stellar metallicity. Indeed, to zeroth order, the result
results of IL04 can be understood by adopting their prescription for
core growth (equation \ref{mct}) and deriving the region of parameter space
of initial conditions, at given Z, that lead to the creation of a 
critical mass core. Wyatt, Clarke \& Greaves (2007) showed that this is
roughly equivalent to thresholding the disc mass distribution at a fixed
mass in solids; they showed that, using the {\it observed} disc mass
distribution (Andrews \& Williams 2005) one can reproduce the
observed dependence of planet frequency on metallicity under this
simple assumption. 

  As noted above, the models of IL04 omitted viscous evolution of the
disc gas and instead depleted the gas on an adjustable global e-folding
timescale. Moreover, they included no hard cut-off in disc gas lifetimes
as would result from photoevaporation. We here therefore revisit this
issue under the assumption that the lifetime of the disc gas is terminated
by (metallicity dependent) photoevaporation. In the context of our
analysis in Section 2, we can readily establish which sets of disc
initial parameters should lead to giant planet formation by looking
at the subset of models for which $t_{form_{min}} < t_{phot}$ (equations
\ref{tphot2}, \ref{tformmin1_p}, \ref{tformmin2_p}). 
We then obtain the criteria:

\begin{equation}
\label{pform1_p}
M_{g0} > K_1 \cdot r_{d0}^{2(2-p)^2/(9-4p)}  Z^{-(5-2p)/(9-4p)} \dot M_W(Z)^{2(5-3p)/5(9-4p)}
\end{equation}

if $a_{ice} > a_{tg}$ and

\begin{equation}
\label{pform2_p}
M_{g0} > K_2 \cdot \bigl(r_{d0}^{-(2-p) (55-11p+4p^2)} Z^{-(13+2p)(5-2p)} \dot M_W(Z)^{2(2-p)(5-3p)} \bigr)^{(1/(105-56p+6p^2)}
\end{equation}

if $a_{ice} < a_{tg}$. $K_1$ and $K_2$ are constants.

We thus obtain the expected result that a higher threshold gas mass is
required at low metallicities; therefore if we make the reasonable
assumption that the disc gas {\it mass} distribution should be independent
of metallicity one recovers the qualitative result that the incidence
of giant planets should increase with metallicity. In order to quantify
this, one of course has to make specific assumptions about the
form of the distributions of initial disc masses and radii (Wyatt
et al. 2007). Here, however, our interest lies elsewhere: we just want
to discover whether the particular aspect of metallicity dependent
photoevaporation rates is an important factor in driving this
relationship. To this end we compare the powers of Z in equations
\ref{pform1_p} and \ref{pform2_p} that would result from the case that
the photoevaporation rate was independent of Z (i.e. $\dot M_W(Z)$ =
constant) with the power of Z in the case that we employ our
X-ray derived value $\dot M_w(Z) \propto Z^{-0.77}$. For $p=1$ we find
that this changes the power of Z in equation \ref{pform1_p} from
$-0.6$ to $-0.88$ and in equation \ref{pform2_p} from
$-0.8$ to $-0.83$. This therefore implies that the metallicity
dependence of the photoevaporation rate has a rather minor part to play
in explaining the observed positive correlation between planet
frequency and stellar metallicity. This can be readily understood
inasmuch as we have already seen (equation \ref{tphot}) that the
disc lifetime is not expected to be a strong function of metallicity
in this case and thus this does not in itself have a large impact
on the planet producing capacity of a disc.

\section{Conclusions}

We have shown that the study of disc lifetimes in regions of different
metallicities can be used as a powerful discriminant between the two
currently leading models of disc dispersal -- photoevaporation and
planet formation. By means of detailed 
thermal and photoionisation calculations we have determined that a
disc's lifetime against photoevaporation is a shallow positive power of
metallicity, $t_{\rm phot} \propto Z^{0.52}$. This metallicity
dependance is specific to a photoevaporation mechanism driven mainly
by X-ray radiation as it relies on the significant reduction of the
gas opacities with metallicity. We have also shown that, on the
contrary, a disc dispersal mechanism based on planet formation yields
disc lifetimes that are a strong negative power of $Z$. Therefore, a census
of disc fractions in regions of lower metallicities compared to the
solar neighbourhood will be crucial to determine which is the dominant
mechanism responsible for the rapid demise of protoplanetary discs. 

Recent observations of embedded clusters in Cloud 2 of the Extreme
Outer Galaxy (EOG) presented by Yasui et al. (2009, YS09) indeed find
shorter disc lifetimes for this metal poor environment ([O/H]$\sim$-0.7 dex),
compared to the solar neighbourhood. These results are in agreement
with the predictions of an X-ray+EUV photoevaporation model, and argue
against planet formation as the dominant dispersal mechanism.  

 YS09 quote a mass detection limit of 0.1~M$_{\odot}$ for Cloud2 in the
EOG using the 8.2m {\it Subaru} telescope, similar to values obtained
for embedded clusters in the solar neighborhood with smaller (2-4m
class) telescopes, and thus argue that a comparison of the disc
fractions in embedded clusters in the EOG with those of clusters in
the solar neighbourhood is justified. Further observations aimed at
determining disc fractions in low metallicity regions of various ages
would be very useful to confirm these important results. 

We finally show that, in the context of giant planet formation in the
core accretion scenario, the effect of metallicity dependent
photoevaporation is to strengthen the conclusion that planet
formation is favoured in high metallicity environments since the
lifetime of the disc against photoevaporation ($t_{\rm phot}$) is a
positive function of Z. This effect, however, only plays a secondary
role: the main process that favours planet formation at high
metallicity is simply the faster core growth in the case of a high
surface density of solids in the disc. 

\section*{Acknowledgments}
We thank the reviewer, Jane Greaves, for helpful suggestions that
improved the clarity of this paper. We also thank Mike Meyers and
Stephane Udry for helpful discussions. BE was supported by a Science
and Technology Facility Council Advanced Fellowship.  
This work was performed using the Darwin Supercomputer
of the University of Cambridge High Performance Computing Service
(http://www.hpc.cam.ac.uk/), provided by Dell Inc. using Strategic
Research Infrastructure Funding from the Higher Education Funding
Council for England.

\appendix
\section{Analytical Derivations}

In order to evaluate the metallicity dependence of giant planet formation, we need to
establish how the background nebula evolves as a function of the initial conditions,
i.e. the initial total disc mass in gas, $M_{g0}$, and the initial
disc outer radius, $r_{d0}$. We assume that the evolution of the disc
gas is governed by viscous 
evolution provided that the accretion rate through the disc exceeds
the photoevaporation rate, $\dot M_W(Z)$, and that at this point the
disc gas is dispersed, thus ruling out subsequent gas giant planet
formation. We start by providing a heuristic derivation of 
the asymptotic evolution of a viscous disc  (see Lynden-Bell \& Pringle
1974, Hartmann et al 1998) in which the kinematic viscosity is a power law
function of radius, i.e.

\begin{equation}
\label{nu}
\nu \propto r^p
\end{equation} 

  We first note that such a disc evolves through a sequence of discs with increasing 
radius, $r_d$, and
decreasing gas mass, $M_g$, such that the total disc angular momentum is to first order conserved (since
the fraction of the total disc angular momentum that is advected onto
the star is small). Thus, since most of the disc's angular momentum resides at large radius, we have:

\begin{equation}
\label{jcon}
M_g r_d^{1/2} \sim M_{g0} r_{d0}^{1/2}
\end{equation}

Furthermore, the disc evolves on the viscous timescale at $r_d$; since this is long compared with
the viscous timescale at radii $<< r_d$, it follows that the disc at radii $<< r_d$ is in
an approximately steady state, i.e. the accretion rate, $\dot M$, is
independent of r. In this case, accretion 
disc theory (e.g. Pringle 1981) relates this (at radii well away from
the disc's inner edge) to the kinematic viscosity and surface density,
$\Sigma_g(r)$, via

\begin{equation}
\label{mdot}
\dot M \sim 3 \pi \Sigma_g \nu
\end{equation}

It thus follows that the asymptotic form of the viscously evolving disc over much of its radial
extent is in this case given by the power law $\Sigma_g \propto r^{-p}$.
If we normalise this power law by the instantaneous values of the disc mass and radius we then obtain

\begin{equation}
\label{sigmar}
\Sigma_g \propto M_g r_d^{-(2-p)} r^{-p}
\end{equation}

Combining this with equation \ref{jcon}, we can eliminate $r_d$ and obtain:

\begin{equation}
\label{sigmarg}
\Sigma_g \propto M_{g0}^{-2(2-p)} r_{d0}^{-(2-p)} M_g^{(5-2p)}r^{-p}
\end{equation}

Now from equations \ref{nu}, \ref{mdot} and \ref{sigmarg} we can write

\begin{equation}
\label{mdot2}
\dot M = - {{d M}\over{dt}} \propto M_{g0}^{-(4-2p)} r_{d0}^{-(2-p)} M_g^{(5-2p)}
\end{equation}

from which we obtain the asymptotic power law scalings:

\begin{equation}
\label{mt}
M_g \propto M_{g0} r_{d0}^{1/2} t^{-1/(4-2p)}
\end{equation}

\begin{equation}
\label{mdott}
\dot M \propto  M_{g0} r_{d0}^{1/2} t^{(-5+2p)/(4-2p)}
\end{equation}

and

\begin{equation}
\label{sigmart}
\Sigma_g \propto M_{g0} r_{d0}^{1/2} r^{-p} t^{(-5+2p)/(4-2p)}
\end{equation}

 If we now define the disc lifetime against dispersal by photoevaporation as the
disc age such that $\dot M = \dot M_W(Z)$, we have:

\begin{equation}
\label{tphot}
t_{phot} \propto \dot M_W(Z)^{(4-2p)/(-5+2p)} M_{g0}^{(4-2p)/(5-2p)} r_{d0}^{(2-p)/(5-2p)}
\end{equation}

 We now consider planet formation in the context of such a viscously evolving gas disc and follow
IL04 by assuming that the growth of the rock core is governed by an equation of the form

\begin{equation}
\label{mct}
{{dM_c}\over{dt}} \propto \Sigma_d(r) r^{-3/5} M_c^{2/3} \Sigma_g^{2/5}
\end{equation}

$\Sigma_d$ is the surface density of rocky planetesimals. We follow IL04 by assuming
that this distribution is decoupled from the evolutionary processes that control the evolution
of $\Sigma_g$. We thus assume that a fixed fraction of the initial solid content of the
disc forms a disc whose radial profile follows the initial  radial profile of the gas
and that $\Sigma_d$ is then constant in time (we thus neglect the depletion in $\Sigma_d$
due to the accretion of planetesimals during core growth; this is acceptable to first order
provided that the core is not close to achieving its isolation mass, an issue that we consider
further
 below). Thus

\begin{equation}
\label{sigmad}
\Sigma_d \propto Z M_{g0} r_{d0}^{-(2-p)} r^{-p}
\end{equation}
 
substituting from \ref{sigmart} and \ref{sigmad} into \ref{mct} we obtain:

\begin{equation}
\label{mct2}
{{dM_c}\over{dt}} \propto Z M_{g0}^{7/5} r_{d0}^{-(9-5p)/5} r^{-(7p+3)/5} M_c^{2/3} t^{-(2/5)(5-2p)/(4-2p)}
\end{equation}

  We conclude from \ref{mct2} that the timescale for the formation of a rocky core of critical
mass for the accretion of a gaseous envelope  then scales with radius and initial disc parameters
according to:

\begin{equation}
\label{tform}
t_{form} \propto Z^{-5(2-p)/(5-3p)} M_{g0}^{-7(2-p)/(5-3p)} r_{d0}^{(2-p)(9-5p)/(5-3p)} r^{(2-p)(7p+3)/(5-3p)}
\end{equation}

 Equation \ref{tform} demonstrates that the formation timescale increases with increasing radius, as expected
given the lower surface density and larger orbital timescale at larger radius. Thus planet formation occurs
first at the minimum radius that is allowed according to two further criteria. First of all, it is
necessary that this radius is at least as large as the ice sublimation radius ($a_{ice}$) since the
sticking efficiency of planetesimals is much lower in regions devoid of solid ice. Secondly, it is necessary
that a critical core mass can form without consuming all the planetesimals in its `feeding zone'. 
This latter is a region whose
width is related to the Hill radius of the core, and thus is a linear function of orbital radius. Thus the
radius ($a_{tg}$) at which a core of critical mass just consumes all the planetesimals in its feeding zone obeys
a scaling of the form

\begin{equation}
\label{atg}
\Sigma_d a_{tg}^2 = {\rm{constant}}
\end{equation}
i.e.
\begin{equation}
\label{atg2}
a_{tg} \propto Z^{-1/(2-p)} r_{d0} M_{g0}^{-1/(2-p)}
\end{equation}

 Therefore if $a_{ice} > a_{tg}$ (i.e. if the minimum radius for planet formation is set by ice sublimation
rather than feeding zone considerations) we have:

\begin{equation}
\label{tformmin1}
t_{form_{min}} \propto \bigl( Z^{-5} M_{g0}^{-7} r_{d0}^{(9-5p)} \bigr)^{(2-p)/(5-3p)}
\end{equation}

Alternatively, if $a_{tg}>a_{ice}$ we substitute \ref{atg2} into \ref{tform} and obtain

\begin{equation}
\label{tfrommin2}
t_{form_{min}} \propto \bigl( Z^{(-13+2p)} M_{g0}^{-17} r_{d0}^{(4-2p)(6+p)} \bigr)^{1/(5-3p)}
\end{equation}

  Finally, the condition that at least one gas giant planet is able to form prior to disc photoevaporation is
given by the condition $t_{form_{min}} < t_{phot}$, i.e.

\begin{equation}
\label{pform1}
M_{g0} > K_1 \cdot r_{d0}^{2(2-p)^2/(9-4p)}  Z^{-(5-2p)/(9-4p)} \dot M_W(Z)^{2(5-3p)/5(9-4p)}
\end{equation}

if $a_{ice} > a_{tg}$ and

\begin{equation}
\label{pform2}
M_{g0} > K_2 \cdot \bigl(r_{d0}^{-(2-p) (55-11p+4p^2)} Z^{-(13+2p)(5-2p)} \dot M_W(Z)^{2(2-p)(5-3p)} \bigr)^{(1/(105-56p+6p^2)}
\end{equation}

if $a_{ice} < a_{tg}$. K$_1$ and K$_2$ are constants.

\end{document}